\documentclass[aps,pre,preprint,superscriptaddress,onecolumn]{revtex4-2}

\usepackage{color}

\usepackage{epsfig,amssymb,amsmath,amsthm,amsfonts,amsbsy,mathrsfs}

\usepackage{graphicx}

\usepackage{subfigure}

\usepackage{float}

\usepackage{verbatim}

\usepackage{physics}

\newcommand{\xb}{\mathbf{x}}
\newcommand{\psid}{\psi_\delta}

\newcommand{\sgn}{{\rm sign}\,}
\newcommand{\Jb}{{\bf J}}
\renewcommand{\i}{{\rm \bf i}}
\renewcommand{\Im}{{\rm Im}\,}

\begin{document}

\title[Nematic alignment of self-propelled rods]{A mean-field model for nematic alignment of self-propelled rods}
\author{Misha Perepelitsa\footnote[1]{These authors contributed equally to this project}
\footnote[2]{corresponding authors}
}
\affiliation{Department of Mathematics, University of Houston, TX 77204}

\author{$\rm Ilya\,Timofeyev^{\textrm{a}}$}
\affiliation{Department of Mathematics, University of Houston, TX 77204}

\author{$\rm Patrick\, Murphy^{\textrm{a}}$}
\affiliation{Department of Bioengineering, Rice University, Houston, TX 77005}

\affiliation{Center for Theoretical Biological Physics, Rice University, Houston, TX 77005}

\author{\; $\rm Oleg\, A.\, Igoshin^{\textrm{b}\,,}$}
\affiliation{Department of Bioengineering, Rice University, Houston, TX 77005}

\affiliation{Center for Theoretical Biological Physics, Rice University, Houston, TX 77005} 

\affiliation{Department of Chemistry, Rice University, Houston, TX 77005}

\affiliation{Department of Biosciences, Rice University, Houston, TX 77005}

\date{\today}

\begin{abstract}
Self-propelled rods are a facet of the field of active matter relevant to many physical systems ranging in scale from shaken granular media and bacterial alignment to the flocking dynamics of animals. In this paper we develop a model for nematic alignment of self-propelled rods interacting through binary collisions. We avoid phenomenological descriptions of rod interaction in favor of rigorously using a set of microscopic-level rules. Under the assumption that each collision results in a small change to a rod's orientation, we derive the Fokker-Planck equation for the evolution of the kinetic density function. Using analytical and numerical methods, we study the emergence of the nematic order from a homogeneous, uniform steady-state of the mean-field equation. We compare the level of orientational noise needed to destabilize this nematic order and compare our results to an existing phenomenological model that does not explicitly account for the physical collisions of rods. We show the presence of an additional geometric factor in our equations reflecting a reduced collisions rate between nearly-aligned rods reduces the level of noise at which nematic order is destroyed, suggesting that alignment  that depends on purely physical collisions is less robust.
\end{abstract}

\maketitle

\section{Introduction}
Interactions between self-propelled rods have been the subject of many studies on what is known as active matter. These studies were motivated by attempts to understand interactions of rod-shaped cells \cite{Kaiser03,Balagam14b,Balagam15b,Zhang10,Be'er20} and  motor-driven filaments such as microtubules \cite{Ndlec97,Schaller10,Sumino12}. How the mechanisms of rod interactions in such systems give rise to self-organization is a fundamental question that continues to draw interest. For example, mechanical interactions between the rods result in their local alignment and, under certain conditions, the emergence of nematic order. These phenomena  have been extensively studied in the spirit of the alignment model of Vicsek et al. \cite{Vicsek1995}, see for example Chat\'{e} et al. \cite{Chate2006}, Peruani et al. \cite{Peruani2008}, Ginelli et al. \cite{Ginelli2010}, Bolley et al. \cite{Bolley12}, and Degond et al. \cite{Degond2017}.  
Such models typically assume the collective alignment of nearby rods through mean nematic current or mean nematic orientation, also known as the director. However, the rules proposed in many of these studies lack first-principle derivations. %The models of this type are convenient for analytical treatment because they result in mean-field type equations that can be analyzed for stability, phase transition, etc. 

%In lower density cases, pairwise collisions between cells are the most common type of interactions. During an interaction, each cell re-orients and continue to move in the new direction.
%An direct approach based on tracking individual cells and implementing modeling of interactions of cell as mechanistic systems of self propelled rigid rods, is possible, \cite{} but is computationally prohibitive for large number of cells, in part due to increasing complexity of interactions involving multiple cells.

Boltzmann models of binary collisions provide a way to rigorously derive the form of the collective alignment in scenarios where the rods interact primarily through physical collisions rather than long range interactions. For example, binary collision models for polar alignment of self-propelled particles were studied by Bertin et al. \cite{Bertin2006}. The authors obtained a Boltzmann-type equation for rod number density which they reduced to a system of macroscopic (hydrodynamic) equations for the first two moments of the number density. The steady-states of the latter system were used to extract information about the phase transition to an ordered phase. In the context of non-polar alignment, Hittmeir et al. \cite{Hittmeir2020} considered a model for the alignment of myxobacteria, deriving a Boltzmann-type equation for the cell motion, a closed system of macroscopic equations, and proving a number of
analytical results about the existence of solutions for these equations. Notably, this model was based on instantaneous alignment and reversal from collisions depending on the difference in orientations, and the rules for collisions were structured to include reversals at the expense of nematic alignment.

%In \cite{Bertin2006, Hittmeir2020} the typical macroscopic scale of motion is equals $L\sim Nl,$ where $N$ is the number of rods and $l$ is the length of the rod, while in our model, it is $L\sim Nl\delta,$ where $\delta$ is the fraction of angle change per collision. 

Motivated by the mechanics of motion of {\it Myxococcus xanthus} \cite{Kaiser03,Munoz16} and monolayers of {\it Bacillus subtilis} \cite{Kearns03,Zhang10,Be'er20}, which mainly reorient through physical-contact interactions, we consider a model for the collective behavior of self-propelled rods in which nematic re-orientation occurs gradually over a series of sequential binary collisions. These can be collisions between the same two rods or with others nearby. Our goal is to derive from microscopic collision rules a tractable continuous-time equation governing the evolution of the rods' spatial distribution, which we can then use to obtain quantitative information about the impact of density and noise parameters on self-organization. We assume that each collision results in the re-orientation of rods by a certain amount, assumed to be small. Thus, the change in a rod's orientation over some time is the result of cumulative small changes due to multiple collisions. The nematic alignment we consider differs from \cite{Hittmeir2020} in this respect.

The paper is organized as follows. In section \ref{sec:collision}, we derive an Enskog-type equation for the rod number density and, in a certain asymptotic regime, the mean-field Fokker-Planck equation. Our approach is similar to the treatment for grazing collisions of molecules in gas dynamics \cite{CercignaniBook}. We introduce an averaging type agent-based model that, at the macroscopic level, is equivalent to the binary collision model (BC model). The Fokker-Planck model obtained for the BC-model turns out to be similar to the heuristic liquid crystal model (LC model) of Peruani et al. \cite{Peruani2008}. Our kinetic equation differs from the latter by a factor in the alignment rate that accounts for a decrease in collisions as cells become nematically aligned. This naturally leads to the question of comparing qualitative and quantitative differences between dynamics generated by these two models. We address this question in section \ref{sec:LC} where we consider the stability of a spatially homogeneous steady state with a uniform distribution of orientations for both alignment models augmented by diffusion (noise) in rod orientation. Linear stability can be calculated explicitly, as was done in \cite{Peruani2008}. In section \ref{sec:NL} we discuss the nonlinear stability for the models of interest, using a numerical solver for the nonlinear mean-field equations. We obtain parameter values for the phase transition to a nematically ordered phase and compute steady-state orientation distributions for several representative cases. Our findings show that self-propelled rods are less ordered in the BC model compared to the LC model due to decreased chance of collisions between cells with similar nematic orientations. Therefore, the transition to the disordered phase occurs at a lower level of rotational noise.

%In addition, our approach results in a formula for the strength of alignment in terms of the microscopic parameters of motion, formula \eqref{kappa}, while in \cite{Peruani2008} it was left unspecified. 

\section{Binary collision model}\label{sec:collision}

\subsection{Derivation of Boltzmann-type equation}
\begin{figure}[h!]
\centering
\includegraphics[width=0.9\textwidth]{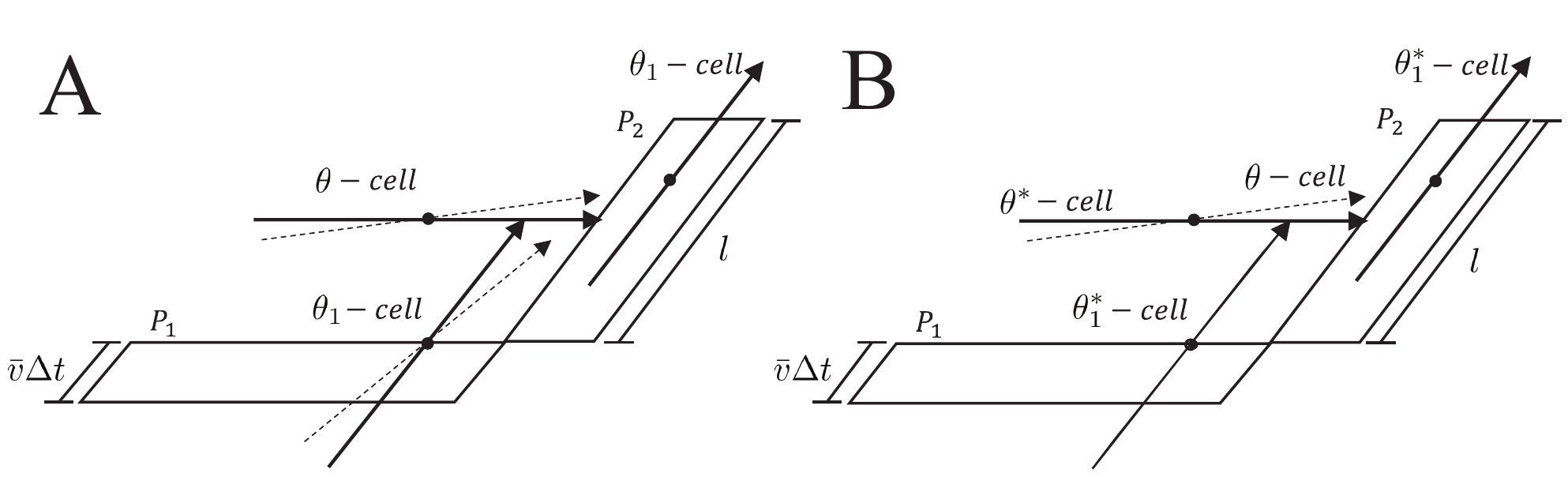}

\caption{Alignment geometry for a colliding rod changing from (A) or to (B) a given orientation $\theta$. A) Parallelogram $P_1$ is formed by vectors $l\mathbf{e}(\theta)$ and $\bar{v}\Delta t\,\mathbf{e}(\theta_1).$
Parallelogram $P_2$ is formed by vectors $l\mathbf{e}(\theta_1)$ and $\bar{v}\Delta t\,\mathbf{e}(\theta).$
A $\theta$--rod (solid arrow) re-orients to a different orientation (dotted arrow) upon interaction with a $\theta_1$-rod from $P_1$ or $P_2.$ $\theta_1$--rod re-orients by the same angle in the opposite direction (dotted arrow). B) Same as A) but with a collision in $P_1$ or $P_2$ resulting in a $\theta$ rod.
\label{fig:collision}
}
\end{figure}

We consider a collection of $N$ rods moving on a square domain $[0,L]\times[0,L]$ with periodic boundary conditions.  Rods are rigid and move along their longer axes, with their midpoint denoted by the coordinates $\mathbf{x}$. Let $l$ be the length of the rod with corresponding unit orientation vector $\mathbf{e}$. We denote the rod speed by $\bar{v}.$ 
In our model we assume that the collision between a rod with orientation $\mathbf{e}=\mathbf{e}(\theta) = (\cos\theta,\sin\theta)$, which we call a $\theta$-rod, with a $\theta_1$-rod results in a nematic re-orientation of both rods, yielding in new angles $\theta^*,$ and $\theta^*_1$ according to the rule
\begin{equation}
\label{rule2.1}
\begin{split}
\theta^* &=\theta + \delta\phi(\theta_1-\theta),\\
\theta_1^* &= \theta_1 -\delta\phi(\theta_1-\theta),
\end{split}
\end{equation}
where $\phi(\theta)$ is a $\pi$--periodic, odd function, positive on the interval $(0,\pi/2)$ and negative on the interval $(\pi/2,\pi),$ and $\delta \in (0,1)$ is small numeric parameter that controls the strength of alignment. We are implicitly assuming here that $\delta$ is the result of rescaling $\phi$ to order $O(1)$.
The function $\phi= \sin(2\theta)$ is a typical example with a microscopic interpretation that we will use in section \ref{sec:LC}. We will assume that a collision between rods occurs when the head of one rod is in contact with the other (see Figure \ref{fig:collision}). We will assume that interactions a predominantly binary and the contribution from simultaneous interaction of more than two rods is negligible. We also allow for rods to overlap each other spatially.  We will refer to this model of alignment as a binary collision (BC) model.  The probabilistic description of alignment is based on the one-particle distribution function of rod positions and orientations, $f,$  with corresponding rod number density $F(\xb,\theta,t) = Nf(\xb,\theta,t)$. As collisions happen between pairs of particles, we will also need the 2-particle distribution function $f^{(2)}(\xb,\theta,\xb_1,\theta_1,t)$ which stands for the density of the joint probability distribution for a random pair of cells, and the corresponding function $F^{(2)}(\xb,\theta,\xb_1,\theta_1,t)=N(N-1)f^{(2)}(\xb,\theta,\xb_1,\theta_1,t)$, which stands for the number of pairs of cells (averaged over an ensemble) when given two spatial regions and two sets of orientations. 

We first must calculate the change in $F = Nf$ due to collisions. The value of $f$ at a given orientation $\theta$ decreases when a $\theta$-rod interacts with a $\theta_1$-rod ($\theta_1\not=\theta$) (Figure \ref{fig:collision}a) and increases when a $\theta^*$-rod and a $\theta^*_1$-rod produce a $\theta$-rod post-collision (Figure \ref{fig:collision}b). To describe the latter we need to swap the roles of $\theta$ and $\theta^*$ in \eqref{rule2.1}, then use the system to find $\theta^*$ in terms of the pre-collision orientation $\theta^*_1$ as well as the given angle $\theta$. First, 
\begin{equation}
\label{rule3.1}
\begin{split}
\theta &= \theta^* + \delta\phi(\theta_1^*-\theta^*).%\\
%\theta_1 &= \theta_1^* -\delta\phi(\theta_1^*-\theta^*).
\end{split}
\end{equation}
It follows using substitution that $\theta^*_1-\theta = \theta^*_1-\theta^* -\delta\phi(\theta^*_1-\theta^*).$ Denote the solution $\theta$ of \[
\omega=\theta-\delta\phi(\theta)
\]
by $\psi_\delta(\omega)$. Then, we compute 
\begin{equation}
\label{eq:inv}
\partial_\omega \psi_\delta= (1-\delta \phi'(\psi_\delta(\omega)))^{-1},
\end{equation}
which shows that the inverse function $\psi_\delta$ exists when $\delta$ is sufficiently small. Given $\theta$ and an arbitrary pre-collisional orientation $\theta_1^*,$ we can determine the other pre-collisional orientation $\theta^*$ so that $(\theta^*,\theta_1^*)$ collision  results in a $\theta$-rod, given by
\begin{equation}
\label{eq:theta'}
\theta^*=\theta_1^*-\psid(\theta_1^*-\theta).
\end{equation}

We now need to calculate the change in the number of rods over the time interval $(t,t+\Delta t)$, denoted by $N\Delta f$. We do this by splitting the change into decreasing and increasing contributions $\Delta f = \Delta^- f - \Delta^+ f$, with $\lim_{\Delta t \to 0} (\Delta^- f - \Delta^+ f)/\Delta t = (\frac{df}{dt})^+ - (\frac{df}{dt})^- = \frac{df}{dt}$, the total or Lagrangian derivative of $f$. This captures the total change in time of the given quantity with respect to all variables that $f$ depends on, and can alternatively be expressed via the chain rule as
\begin{equation}
    \frac{df}{dt} = \partial_t f + \sum_{i=1}^2 \partial_{x_i}f\, \frac{d x_i}{dt} + \partial_\theta f \,\frac{d\theta}{dt}.
\end{equation}
In our context we have $\frac{d\theta}{dt} = 0$ and $(\frac{d x_1}{dt},\frac{d x_2}{dt}) = \bar{v}(\cos(\theta),\sin(\theta))$
due to our assumptions about the self-propelled nature of the rods.

First, we approximate the number of rods within a given spatial region and orientation range by $Nf(\xb,\theta,t)d\xb \, d\theta$. Figure \ref{fig:collision}a shows the geometry involved when calculating possible collisions corresponding to loss with $P_1$ and $P_2$ as the parallelograms where collisions can occur with a given rod. Figure \ref{fig:collision}b corresponds similarly to gain of cells with a given orientation with parallelograms $P_1$ and $P_2$. The magnitude of decrease in the number of rods with orientation angle in $(\theta, \theta+d\theta)$ in $\Delta t$ time equals
the number of rods in this range times the number of collisions, per rod, with another rod in $(\theta_1, \theta_1+d\theta_1)$. Since the differentials $d\xb$ and $d\theta$ ultimately drop out, we do keep track of them for these steps. The expected number of collisions with rods in this range is given by summing the expected number of pairs where the second rod is in $P_1$ or $P_2$. This is made formal by integrating the pairwise joint density $N(N-1)f_2$ over these domains and over $\theta_1$.
\begin{equation}
\label{eq:delta+}
N\Delta^-f(\xb,\theta,t){}={}N(N-1)\left(\int_{-\pi}^\pi\int_{P_1}f^{(2)}(\xb,\theta,\xb_1,\theta_1,t)\,d\xb_1d\theta_1{}+{}\int_{-\pi}^\pi\int_{P_2}f^{(2)}(\xb,\theta,\xb_1,\theta_1,t)\,d\xb_1 d\theta_1\right),
\end{equation}
where the parallelograms $P_1$ and $P_2$ are as in Figure \ref{fig:collision}. 

%To avoid an analogue of the Bogoliubov–Born–Green–Kirkwood–Yvon (BBGKY) hierarchy \cite{harris-book} and 
To obtain a closed-form equation for $f$, we assume the equivalent of molecular chaos, expressed as statistical independence of the pairwise joint distribution $f^{(2)}$ in terms of the single particle distribution
\begin{equation}
\label{eq:indep}
    f^{(2)}(\xb,\theta,\xb_1,\theta_1,t) = f(\xb,\theta,t)f(\xb_1,\theta_1,t).
\end{equation}
This yields 
\begin{equation}
\label{eq:delta+2}
\Delta^-f(\xb,\theta,t){}={}(N-1)\int_{-\pi}^\pi\left(\int_{P_1}f(\xb,\theta,t)f(\xb_1,\theta_1,t)\,d\xb_1{}+{}\int_{P_2}f(\xb,\theta,t) f(\xb_1,\theta_1,t)\,d\xb_1\right)d\theta_1.
\end{equation}
A change of variables from Cartesian $\xb = (x_1, x_2)$ to an $\mathbf{e}(\theta)$, $\mathbf{e}(\theta_1)$ coordinate system gives an extra factor of $|\sin(\theta_1-\theta)|$ in the integrand, taking into account the relative nematic orientations when determining the collision frequency of two rods with orientations $\theta$ and $\theta_1$. This is analogous to the collision factor derived in \cite{Weber13}. Since the length of integration in the $\mathbf{e}(\theta)$ direction is proportional to $\bar{v}\Delta t$, by dividing both sides of \eqref{eq:delta+} by $\Delta t$ and taking the limit $\Delta t \to 0$, we obtain the magnitude of the negative contribution to the rate of change of $f(\xb,\theta,t)$ over all possible $\theta_1$ as
\begin{multline}
\bigg(\frac{df}{dt}\bigg)^-{}={}\bar{v}(N-1)\int_{-\pi}^\pi\int_{-l/2}^{l/2}|\sin(\theta_1-\theta)|f(\xb,\theta,t)\left( f(\xb-\frac{l}{2}\mathbf{e}(\theta_1)+z\mathbf{e}(\theta),\theta_1,t) \right.\\
\left. {}+{}f(\xb+\frac{l}{2}\mathbf{e}(\theta)+z\mathbf{e}(\theta_1),\theta_1,t)\right)\,dz d\theta_1,
\end{multline}
where $z$ is the variable of integration in the $\mathbf{e}(\theta_1)$-direction.

Similarly, the increase $N\Delta^+f$ can be expressed using the expected number of collisions for a rod in $(\theta^*,\theta^*+d\theta^*)$ with a $(\theta_1^*,\theta_1^*+d\theta_1^*)$-rod in $\Delta t$ time such that the result is a rod in $(\theta,\theta+d\theta)$. For this derivation, we need to proceed carefully, taking into consideration the relationship between $\theta$, $\theta^*$, and $\theta_1^*$. In particular we need use the mapping \eqref{eq:theta'} between $\theta$ and $\theta^*$ when relating $f(\xb,\theta,t)d\xb \, d\theta$ to $(f(\xb,\theta^*,t)d\xb \,d\theta^*)(f(\xb_1,\theta_1^*,t)d\theta_1^*)$. Since the differential $d\xb$ again drops out by the end, we do keep track of it. 
As in the previous case, there are two types of collisions depending on whether $\theta_1^*\in P_1$ or $\theta_1^*\in P_2$ (see Figure \ref{fig:collision}b): 
\begin{multline}
\label{eq:plusterm}
\Delta^+ f(\xb,\theta,t)d\theta{}={}-(N-1)\int_{-\pi}^\pi\bigg(\int_{P_1}f(\xb,\theta^*,t)d\theta^* f(\xb_1,\theta_1^*,t)\,d\xb_1 \\ 
{}+{} \int_{P_2}f(\xb,\theta^*,t)d\theta^* f(\xb_1,\theta_1^*,t)\,d\xb_1\bigg)d\theta_1^*.
\end{multline}
Using \eqref{eq:theta'} and \eqref{eq:inv} at $\omega = \theta_1^*-\theta$, we find
\begin{equation}
\label{eq:dtheta'}
d\theta^*{}={} \psi'_\delta(\theta_1^*-\theta)d\theta = (1-\delta\phi'(\psid(\theta_1^*-\theta)))^{-1}d\theta.
\end{equation}
Note that $\theta$ and $\theta_1^*$ are independent, so $d\theta$ in \eqref{eq:plusterm} subsequently drops out. We then obtain the positive rate term in the limit $\Delta t \to 0$ as
\begin{multline}
\bigg(\frac{df}{dt}\bigg)^+{}={}\bar{v}(N-1)\int_{-\pi}^{\pi}\int_{-l/2}^{l/2}\psi'_\delta(\theta_1^*-\theta)|\sin(\psid(\theta_1^*-\theta))|f(\xb,\theta^*,t)\times \\
\bigg(f(\xb-\frac{l}{2}\mathbf{e}(\theta_1^*) + z\mathbf{e}(\theta^*),\theta_1^*,t){}+{}f(\xb+\frac{l}{2}\mathbf{e}(\theta^*)+z\mathbf{e}(\theta_1^*),\theta_1^*,t)\bigg)\,dzd\theta_1^*.
\end{multline}
% with $\theta^*$ given by \eqref{eq:theta'}.

Finally, using that the total derivative of $f$ can be expressed as $\partial_t f + \bar{v}\mathbf{e}\cdot\grad_\xb f$, we get 
the kinetic equation
\begin{multline}
\partial_t f{}+{}\bar{v}\mathbf{e}(\theta)\cdot\grad_\xb f\\
{}={}(N-1)\bar{v}\int_{-\pi}^{\pi}\int_{-l/2}^{l/2}\psi'_\delta(\theta_1^*-\theta)|\sin(\psid(\theta_1^*-\theta))|f(\xb,\theta^*,t)\left(
f(\xb-\frac{l}{2}\mathbf{e}(\theta_1^*) + z\mathbf{e}(\theta^*),\theta_1^*,t)\right. \\
\left.
{}+{}f(\xb+\frac{l}{2}\mathbf{e}(\theta^*) + z\mathbf{e}(\theta_1^*),\theta_1^*,t)\right)\,dzd\theta_1^*\\
-(N-1)\bar{v}\int_{-\pi}^\pi\int_{-l/2}^{l/2}|\sin(\theta_1^*-\theta)|f(\xb,\theta,t)\left( f(\xb-\frac{l}{2}\mathbf{e}(\theta_1^*)+z\mathbf{e}(\theta),\theta_1^*,t) \right.\\
\left. {}+{}f(\xb+\frac{l}{2}\mathbf{e}(\theta)+z\mathbf{e}(\theta_1),\theta_1^*,t)\right)\,dz d\theta_1^*.
\end{multline}
Note that we also changed the notation from $\theta_1$ to $\theta_1^*$ in the last integral to match variables of integration. We will take care of the remaining occurrences of  $\theta^*$ in the next section's asymptotics.

\subsection{Asymptotic expansion and mean-field derivation}
Let $T=\frac{L}{\bar{v}}$ be the characteristic time scale associated with traversing the domain length $L.$ We scale the variables by
\begin{equation}\label{eqScale}
\tilde{\xb}=\frac{\xb}{L},\quad \tilde{t}{}={}\frac{t}{T},\quad \tilde{f}(\tilde{\xb},\theta, \tilde{t}){}={}L^2f(\xb,\theta,t). 
\end{equation}
The kinetic equation scales accordingly and takes the following form (with a slight abuse of notation, we keep using the old notation $\xb,t,f$ for the new variables $\tilde{\xb},\tilde{t},\tilde{f}).$ 
\begin{multline}
\label{eq:kinetic}
\partial_t f{}+{}\mathbf{e}(\theta)\cdot\grad_\xb f\\
{}={}(N-1)\int_{-\pi}^{\pi}\int_{-l/(2L)}^{l/(2L)}\psi'_\delta(\theta_1^*-\theta)|\sin(\psid(\theta_1^*-\theta))|f(\xb,\theta^*,t)\left(
f(\xb-\frac{l}{2L}\mathbf{e}(\theta_1^*) + z\mathbf{e}(\theta^*),\theta_1^*,t)\right. \\
\left.
{}+{}f(\xb+\frac{l}{2L}\mathbf{e}(\theta^*) + z\mathbf{e}(\theta_1^*),\theta_1^*,t)\right)\,dzd\theta_1^*\\
-(N-1)\int_{-\pi}^\pi\int_{-l/(2L)}^{l/(2L)}|\sin(\theta_1^*-\theta)|f(\xb,\theta,t)\left( f(\xb-\frac{l}{2L}\mathbf{e}(\theta_1^*)+z\mathbf{e}(\theta),\theta_1^*,t) \right.\\
\left. {}+{}f(\xb+\frac{l}{2L}\mathbf{e}(\theta)+z\mathbf{e}(\theta_1),\theta_1^*,t)\right)\,dz d\theta_1^*.
\end{multline}
Assuming small $\delta$, we introduce the series approximations
\begin{eqnarray}
\theta^*{}&=&{}\theta-\phi(\theta_1^*-\theta)\delta+O(\delta^2),\\
\psi'_\delta(\theta_1^*-\theta){}&=&{}1 +\phi'(\theta_1^*-\theta)\delta +O(\delta^2), \\
\psid(\theta_1^*-\theta){}&=&{}(\theta_1^*-\theta) +\phi(\theta_1^*-\theta)\delta +O(\delta^2),\\
f(\xb,\theta^*,t)-f(\xb,\theta,t) {}&=&{} \partial_\theta f(\xb,\theta,t)(\theta^*-\theta) + O((\theta^*-\theta)^2)\\
{}&=&{}-\phi(\theta_1^*-\theta)\partial_\theta f(\xb,\theta,t)\delta + O(\delta^2). \nonumber
\end{eqnarray}
Additionally, for any continuous function $G(\theta)$, we can expand
\begin{multline}
\int_{-\pi}^{\pi}|\sin (\psid(\theta_1^*-\theta))|G(\theta_1^*)\,d\theta_1^*\\
{}={}
\int_{-\pi}^{\pi}\left(|\sin(\theta_1^*-\theta)|+\sgn(\sin(\theta_1^*-\theta))\cos(\theta_1^*-\theta)\phi(\theta_1^*-\theta)\delta\right)G(\theta_1^*)\,d\theta_1^*\\
{}+{}O(\delta^2)\sup|G(\theta)|.
\end{multline}
Finally, we can expand $f$ in powers of $l/L$ to get
\begin{equation}
f(\xb-\frac{l}{2L}\mathbf{e}(\theta_1^*) + z\mathbf{e}(\theta^*),\theta_1^*,t) =f(\xb,\theta_1^*,t) + O(lL^{-1}).
\end{equation}
Substituting all these expansions into the kinetic equation \eqref{eq:kinetic}, we derive  the asymptotic relation
\begin{multline}
\label{eq:ka}
\partial_t f {}+{}\mathbf{e}(\theta)\cdot\grad_\xb f
{}={}-\frac{(N-1)l\delta}{L}\partial_\theta\left( f(\xb,\theta,t)\int_{-\pi}^{\pi}|\sin(\theta_1^* - \theta)|\phi(\theta_1^*-\theta)f(\xb,\theta_1^*,t)\,d\theta_1^*\right)\\
{}+{}O\left(NlL^{-1}\delta^2\right){}+{}O\left(Nl^2L^{-2}\delta\right).
\end{multline}

To finalize the model, we postulate two hypotheses. The first is
that the strength of alignment per collision, $\delta,$ is small and that   $Nl\delta{L^{-1}}$ is non-vanishing and finite. Since $Nl\delta{L^{-1}}=(NL^{-2})(Ll)\delta,$ which can be interpreted as the average number of collisions when traveling the domain length multiplied by the maximum change per collision, this condition implies that many small changes in a rod's orientation accumulate to a finite magnitude. In this case $\delta$ is inversely proportional to the number of rods in a narrow band of width $l$. The second hypothesis is a macroscopic limit assumption: $l{L}^{-1}$ is small and the number of rods, $N,$ is large. Under these hypotheses, the leading order approximation of \eqref{eq:ka} is a mean-field, Fokker-Planck equation
 \begin{equation}
\label{FK1.1}
\partial_t f {}+{}\mathbf{e}(\theta)\cdot\grad_\xb f
{}+{}\kappa\partial_\theta\left( f(\xb,\theta,t)\int_{-\pi}^{\pi}|\sin(\theta_1 - \theta)|\phi(\theta_1-\theta)f(\xb,\theta_1,t)\,d\theta_1\right){}={}0,
\end{equation}
 where 
 \begin{equation}
 \label{kappa}
 \kappa = \frac{Nl\delta}{L},
 \end{equation}
$\delta, \, l/L \ll 1$ and $N\gg1.$
Note that unlike what one might expect, $\kappa$ is not invariant to parameter changes that keep the mean density $N/L^2$ fixed. This is due to the time rescaling being proportional to $L$. If we double $L$ and keep the mean density fixed, $\kappa$ doubles as well, but this is a reflection of one unit of time, and the average number of collisions a rod would undergo in that time interval, also doubling. However, our phase transition results in the next sections are independent of any scaling that keep the mean density fixed.

In the derivation of the kinetic equation we assumed that $\delta$ is small while $\frac{Nl\delta}{L}$ is finite. We would like to describe two model scenarios in which these assumptions can be justified. The first situation is a statistical black box approach.
Suppose that one has a limited information about the precise mechanism of alignment of self-propelled rods, except that a: alignment is nematic and collisions are mostly binary; and b: the change of angles in collisions is proportional to the angle between interacting cells. Suppose that statistical analysis of an experiment reveals
that there is some characteristic length $L$ such that when the average angle between cell orientations is $\theta_0,$
the changes of cell orientation over distance $L$ are proportional to $\theta_0.$ Estimating the average number of collisions over distance $L$ as
$\frac{Nl}{L},$ one can formulate an effective law that each binary interaction changes a cell angle by the amount $\delta\theta_0,$ so that
\[
\frac{Nl}{L}\delta\theta_0\approx \theta_0,
\]
which leads to the smallness of $\delta$ when $Nl/L\gg 1,$ i.e. the number of collisions per length $L$ is large. 
The other situation, perhaps less applicable to the motion of myxobacteria but still of sufficient interest, is the case when the time of the contact of cells in collisions is short and the torque generated during the collision is finite. For example, bacteria could be able to crawl over one another without fully aligning. 
In this case, $\delta$ can be taken proportional to the time of interaction, and $L$ to be the distance over which changes in orientations are of finite magnitude.

\subsection{Alignment to mean nematic orientation}
In what follows, we replace the factor $N-1$ in \eqref{kappa} with $N$ since both have the same asymptotic behavior for $N\gg 1.$ Following \cite{Peruani2008}, we choose the alignment function to be $\phi(\theta) = \sin(2\theta)$. This functional form is commonly used and is related to frictional force balance, commonly found in environments at low Reynolds' number that bacterial cells experience. It can be derived in different contexts, such as bacterial alignment in a colony due to growth \cite{Dell2018}. In our case it comes from the force a colliding rod receives perpendicular to its orientation, proportional to $\sin(\theta)\cos(\theta)$, from opposing frictional forces generated by the second rod.
 
With this choice, one can introduce a local mean nematic orientation and  \eqref{FK1.1} can be interpreted as a continuous alignment to such mean orientation.
The mean orientation is, in general, is specific for each rod, i.e. it depends on  $\theta$  through the rate of collisions with other rods. Indeed, we can write
\begin{multline}
\int_{-\pi}^{\pi}|\sin(\theta_1 - \theta)|\phi(\theta_1-\theta)f(\xb,\theta_1,t)\,d\theta_1{}={}
\Im \int_{-\pi}^{\pi}|\sin(\theta_1 - \theta)|e^{2i(\theta_1-\theta)}f(\xb,\theta_1,t)\,d\theta_1\\
{}={}\Im \left[e^{-2i\theta} \int_{-\pi}^{\pi}|\sin(\theta_1 - \theta)|e^{2i\theta_1}f(\xb,\theta_1,t)\,d\theta_1\right]
{}={}|\Jb|\sin(2(\Theta-\theta)),
\end{multline}
where
\begin{equation}
\Jb(\xb,\theta,t) = \int_{-\pi}^{\pi}|\sin(\theta_1 - \theta)|e^{2i\theta_1}f(\xb,\theta_1,t)\,d\theta_1,
\end{equation}
is a mean nematic current, with  polar angle $2\Theta(\xb,\theta,t)$ defined by the condition:
\begin{equation}
e^{2i\Theta} = |\Jb|^{-1}\int_{-\pi}^{\pi}|\sin(\theta_1 - \theta)|e^{2i\theta_1}f(\xb,\theta_1,t)\,d\theta_1,
\end{equation}
or equivalently as an angle such that 
\begin{equation}
\int_{-\pi}^{\pi}|\sin(\theta - \theta_1)|\sin(2(\theta_1-\Theta))f(\xb,\theta_1,t)\,d\theta_1{}={}0.
\end{equation}
The kinetic equation in this case is
\begin{equation}
\label{FK1.3}
\partial_t f {}+{}\mathbf{e}(\theta)\cdot\grad_\xb f
{}+{}\kappa\partial_\theta\left( f |\Jb|\sin(2(\Theta-\theta))\right){}={}0.
\end{equation}
The discussion of similar models based on the mean nematic current can be found in Chat\'{e} et al. \cite{Chate2006}, Ginelli et al. \cite{Ginelli2010} and Degond et al. \cite{Degond2017}.

\section{Agent-based models of alignment}
\label{sec:ABM} 
% \subsection{Strength of alignment coefficient}
In this section we describe a simplified agent-based model that can be used in Monte Carlo simulations to obtain suitable approximate solutions of \eqref{FK1.1}. In this model we do not need to trace pairwise interactions of rods, but instead use a more computationally efficient method of accumulating alignments over a neighborhood of a rod. While this is the approach often take for phenomenological models of alignment, we use the form of the mean nematic current we derived from microscopic rules in previous sections to help ensure consistency.
Consider an agent-based model of $N$ rods moving on square domain with side length $L$ with periodic boundary conditions. Let the positions of rod centers $\{\xb_i\}$ and orientation angles $\{\theta_i\}$  evolve according to the ODEs
\begin{equation}
\label{mod:LC}
\begin{split}
\frac{d\xb_i}{dt} &= \bar{v}{\bf e}(\theta_i),\\
\frac{d\theta_i}{dt} & = \gamma_a \sum_{|\xb_i-\xb_j|\leq R}|\sin(\theta_j-\theta_i)|\sin(2(\theta_j-\theta_i)) + \sqrt{2D}\xi_i(t),\quad i=1..N,
\end{split}
\end{equation}
where $\gamma_a>0$ is the strength of alignment,  $R$ is the interaction radius,  $D$ is the rotational diffusivity (diffusion coefficient), and $\{\xi_i(t)\}_{i=1..N}$ is the vector of independent whites noises. The above model is closely related to a model of nematic alignment of self-propelled particles by Peruani et al. \cite{Peruani2008} that was built on the geometry of alignment of liquid crystals. In particular it assumes only angular diffusion so that rods move with constant velocity. We will refer to the model of Peruani et al. \cite{Peruani2008} as the LC-model of alignment. The difference between the two models is a factor $|\sin(\theta_j-\theta_i)|$ in the rate of alignment.

We would like to determine ranges of values of $\gamma_a$ and $R$ such that the agent-based model \eqref{mod:LC} corresponds to BC-model from the last section. To this end we compare \eqref{FK1.1} to the the mean-field equation describing the dynamics of the probability density distribution function $f$ for model \eqref{mod:LC}. Following \cite{Peruani2008},  the latter can be expressed  as a Fokker-Plank equation
\begin{equation}
\label{FK:2}
\partial_t f{}+{}\bar{v}\mathbf{e}(\theta)\cdot\grad_\xb f\\
{}+{}\gamma_a N\partial_\theta\left( V_R(\theta)f(\xb,\theta,t)\right) {}={}D\partial_{\theta}^2f,
\end{equation}
with angular transport velocity 
\[
V_R(\theta){}={}\int_{|\xb_1-\xb|<R}\int_{-\pi}^\pi|\sin(\theta_1-\theta)|\sin(2(\theta_1-\theta))f(\xb_1,\theta_1,t)\,d\theta_1 d\xb_1,
\]
found by rewriting the function underlying rate of change in angle due to collisions $\sum_{|\xb_i-\xb_j|\leq R}|\sin(\theta_j-\theta_i)|\sin(2(\theta_j-\theta_i))$ in (\ref{mod:LC}) as a local integral with the number density $F = N f$, then factoring out $N$. $D>0$ is the rotational diffusion coefficient from the rotational noise in the agent based model.

%Note that this is the same as treating the function %$\sum_{|\xb_i-\xb_j|\leq %R}|\sin(\theta_j-\theta_i)|\sin(2(\theta_j-\theta_i))$ as %the derivative in $\theta_i$ of some potential function %$U(\theta_i,x_i)$ that prescribes the rate of alignment, %and that the bounds of integration for $V_R$ match the %bounds for the sum in (\ref{mod:LC}).

Finally, rescaling $x$, $t$, and $f$ with (\ref{eqScale}) yields 
\begin{equation}
\label{eq:BC}
\partial_t f{}+{}\mathbf{e}(\theta)\cdot\grad_\xb f\\
{}+{}\frac{\gamma_a NL}{\bar{v}}\partial_\theta\left(V_{R/L}(\theta)f(\xb,\theta,t)\right) {}={}\frac{DL}{\bar{v}}\partial_{\theta}^2f.
\end{equation}
The additional local integration over a region $|\xb_1-\xb|<\frac{R}{L}$ in $V_{R/L}$ differs from our model \eqref{FK1.1}, but they can be matched with a simple series expansion for $f(\xb,\theta,t)$. Assuming $\frac{R}{L} \ll 1$, we can expand $f(\xb_1,\theta,t)$ in a power series $f(\xb,\theta,t)+O(R/L)$ around $\xb$ analogously to section \ref{sec:collision}, yielding to first order 
\begin{multline}
\label{eq:BC2}
\partial_t f{}+{}\mathbf{e}(\theta)\cdot\grad_\xb f\\
{}+{}\frac{\gamma_a NL}{\bar{v}}\frac{\pi R^2}{L^2}\partial_\theta\left( f(\xb,\theta,t)\int_{-\pi}^\pi|\sin(\theta_1-\theta)|\sin(2(\theta_1-\theta))f_0(\xb,\theta_1,t)\,d\theta_1 \right) = \frac{DL}{\bar{v}}\partial_{\theta}^2f.
\end{multline}
Note that in terms of parameters, \eqref{FK1.1} corresponds to model \eqref{eq:BC2} when $\frac{R}{L} \ll 1$ 
in such a way that $NR^2L^{-2}$ remains constant
and 
\begin{equation}
\kappa {}={}\pi\frac{\gamma_a N}{\bar{v}}\frac{R^2}{L},
\end{equation}
from which we find that the rate of alignment $\gamma_a$ and interaction radius $R$  are functionally dependent:
\begin{equation}
\gamma_aR^2{}={}\frac{l\bar{v}\delta}{\pi}.
\end{equation}

%%%%%%%%%%%%%%%%%%%%%%%%%%%%%%%%%%%%%%%%%%%%%%%%%%%%%
%%%%%%%%%%%%%%%%%%%%%%%%%%%%%%%%%%%%%%%%%%%%%%%%%%%%%%
%\section{Comparison with a ``liquid crystal'' model of myxo alignment}
\section{Stability of a homogeneous uniform steady state}
\label{sec:LC}
\subsection{Linear stability}
Consider the mean-field equation \eqref{FK1.1} with added diffusion in orientation:
 \begin{equation}
\label{FK1.4}
\partial_t f {}+{}\mathbf{e}(\theta)\cdot\grad_\xb f
{}+{}\kappa\partial_\theta\left( f(\xb,\theta,t)\int_{-\pi}^{\pi}|\sin(\theta_1 - \theta)|\phi(\theta_1-\theta)f(\xb,\theta_1,t)\,d\theta_1\right)\\
{}={}\mu\partial^2_\theta f,
\end{equation}
where $\mu>0$ is a non-dimensional diffusion coefficient.
Following \cite{Peruani2008}, we consider the problem of stability in the space of spatially homogeneous solutions $f(\xb,\theta,t)=g(\theta,t).$ The BC-model's equation for $g$ is
\begin{equation}
\label{eq:BCg}
\partial_t g{}+{}\kappa\partial_\theta\left( g\int_{-\pi}^\pi|\sin(\theta_1-\theta)|\sin(2(\theta_1-\theta))g(\theta_1,t)\,d\theta_1 \right)
{}={}\mu\partial_{\theta}^2g.
\end{equation}
The function $g_0 = 1/(2\pi)$ is a steady-state solution of \eqref{eq:BCg}. By introducing the perturbation from steady state $g = g_0+\epsilon g_1$ and by linearizing equation \eqref{eq:BCg} around $g_0,$ we obtain
\begin{equation}
\label{eq:BCgl}
\partial_t g_1{}+{}\frac{1}{2\pi}\partial_\theta\left( \int_{-\pi}^\pi|\sin(\theta_1-\theta)|\sin(2(\theta_1-\theta))g_1(\theta_1,t)\,d\theta_1 \right)
{}={}\eta\partial_{\theta}^2g_1,
\end{equation}
with the rescaled viscosity coefficient $\eta = \frac{\mu}{\kappa}.$ Equation \eqref{eq:BCgl} admits solutions in the form
\begin{equation}
g_1 = e^{\lambda t}e^{i n\theta},\quad n\in \mathbb{Z},
\end{equation}
provided that 
\begin{equation}
\lambda {}={}\frac{\alpha_n}{\pi}-\eta n^2,
\end{equation}
where $\i\alpha_n$ is the eigenvalue corresponding to the eigenfunction $e^{i n \theta}$ of the operator  $I[g]{}={}\int_{-\pi}^\pi|\sin(\theta_1-\theta)|\sin(2(\theta_1-\theta))g(\theta_1,t)\,d\theta_1.$    For all odd $n,$ $\alpha_n$=0, and for even $n$,
\begin{equation}
\alpha_n{}={}2n\left(\frac{1}{(n-1)(n+1)}-\frac{1}{(n-3)(n+3)}\right). 
\end{equation}
Among these values, only $\alpha_2=32/15$ is positive. Thus, when $\eta<\eta_c=\frac{8}{15\pi},$ perturbations of the constant steady state $g_0$  will grow in the direction of $\sin(2(\theta-\theta_0)),$ for some $\theta_0.$ 
For comparison, see \cite{Peruani2008}, we recall that for the LC-model the critical value is $\eta_c = 0.25.$ 
Equation \eqref{eq:BCg} can be used to show that asymptotically, for $\eta$ close to $\eta_c,$ the steady-state amplitude of perturbations is of order $(\eta_c-\eta)^{1/2},$ i.e.
\begin{equation}
g(\theta) \approx g_0(\theta) + c_0(\eta_c-\eta)^{\frac{1}{2}}\sin(2(\theta-\theta_0)).
\end{equation}
The stability condition $\eta< \eta_c$ can be written in terms of parameters of the agent-based model as
a balance between the diffusivity, the density, and the strength of alignment:
\begin{equation}
D > \frac{8}{15}\frac{N}{L^2}(\gamma_a R^2).
\end{equation}
In the linear case the instabilities grow without bound. The nonlinear equation however will take the perturbations to a well defined steady state, which we compute by solving \eqref{eq:BCg} numerically in section \ref{sec:NL}.

\subsection{Nonlinear stability}
\label{sec:NL}
\subsubsection{Numerical scheme}

Equation \eqref{eq:BCg} can be schematically written as
\begin{equation}
\partial_t g + \partial_\theta \left( \mathbf{e} g \right) = \eta\partial_{\theta}^2 g,
\end{equation}
where the velocity $\mathbf{e} \equiv \mathbf{e}(\theta,g,t)$ is computed using 
a nonlocal integral operator. We use an operator-splitting method to numerically integrate equations of his type. Thus, the advection and the diffusion parts of this equation are integrated using different discretizations. The diffusive part is integrated numerically using a standard central finite-difference scheme. We use the local Lax-Friedrichs  method from LeVeque \cite{leveque-book}
for the advection part. In particular, if we consider the wave equation written in flux form $\partial_t g + \partial_\theta F = 0$ (where $F \equiv \mathbf{e}g$), then we use following numerical method to advance the solution during one time-step
\begin{equation}
g_k^{j+1}=g_k^j - \frac{\Delta t}{\Delta \theta} [F_{k+1/2}-F_{k-1/2}]
\end{equation}
where $g_k^j = g(k\Delta \theta, j\Delta t)$ and 
numerical fluxes are given by
\begin{equation}
  F_{k+1/2}= \frac12 (v_k g_k^j + v_{k+1} g_{k+1}^j) - \lambda_{i+1/2}(g_{k+1}^j - g_k^j)
\end{equation}
with $v_k = v(k\Delta \theta, j\Delta t)$ and
\begin{equation}
  \lambda_{i+1/2}=\max(|v_k|,|v_{k+1}|)/2.
\end{equation}
We use $N=400$ points for the discretization and 
$\Delta t=2.5 \times 10^{-3}$ in all simulations. We also verified our simulations with 
$N=800$ and $\Delta t=6.25 \times 10^{-4}$ for several selected simulations (several different initial conditions) and did not observe any significant difference between simulations with low and high resolutions. Thus, we can conclude that simulations with
 $N=400$ and $\Delta t=2.5 \times 10^{-3}$ are well-resolved and the numerical diffusion due to the Lax-Friedrichs discretization of the advection part is not significant.

\subsubsection{Numerical simulations}

\begin{figure}[htbp]
\centering
\includegraphics[clip, trim=0.5cm 8.5cm 0.5cm 9.2cm, width=1.00\textwidth]{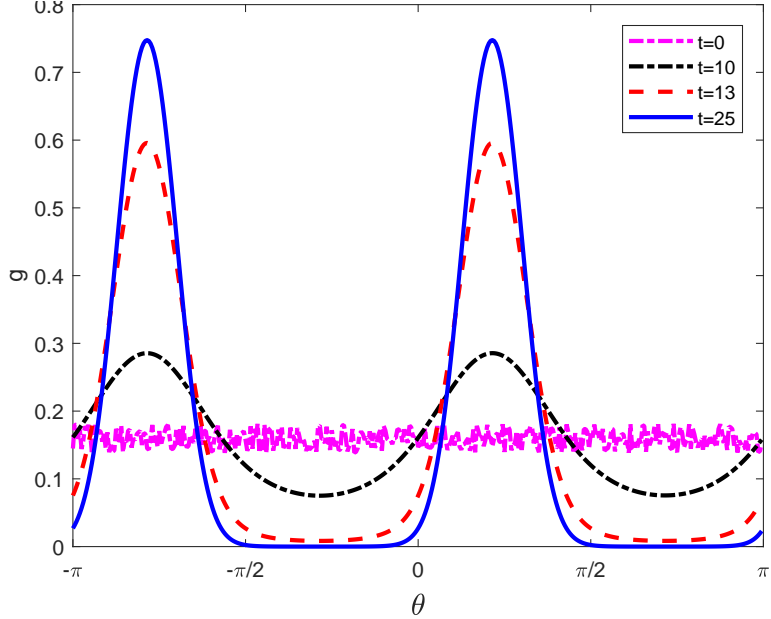}

\caption{Evolution of $g(\theta,t)$ to a steady state in ustable regime. The result of numerical simulation of equation \eqref{eq:BCg} at times $t=0,\; 10,\; 13,\; 25$ starting from random initial data. By time $t=25$ the numerical solution reaches the steady state.    
\label{fig:cuts}
}
\end{figure} 

\begin{figure}[htbp]
\centering
\centerline{\includegraphics[clip, trim=0.5cm 8.5cm 0.5cm 8.7cm, width=0.7\textwidth]{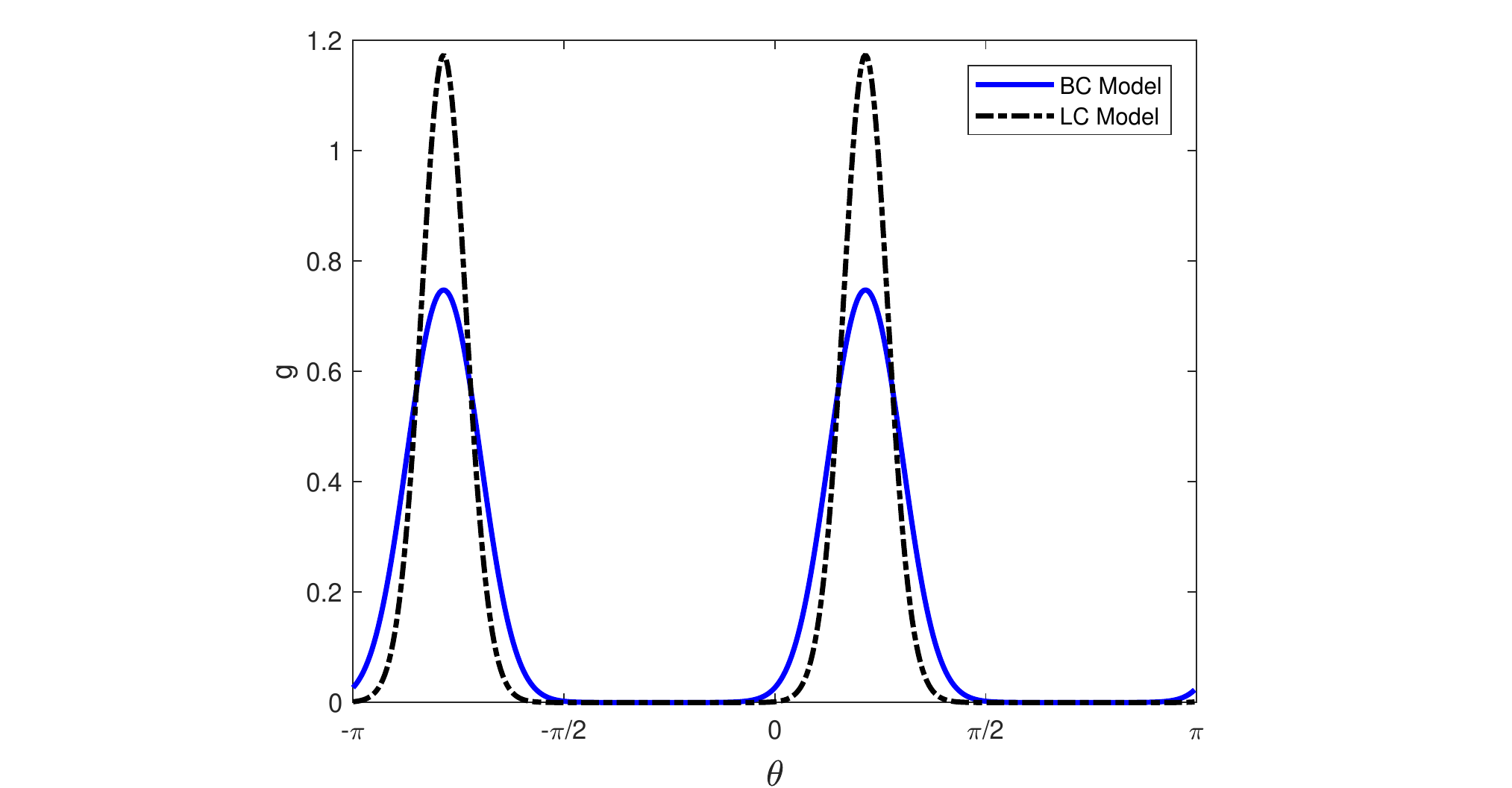}\hspace*{-3.0cm}
\includegraphics[clip, trim=0.5cm 8.5cm 0.5cm 8.7cm, width=0.7\textwidth]{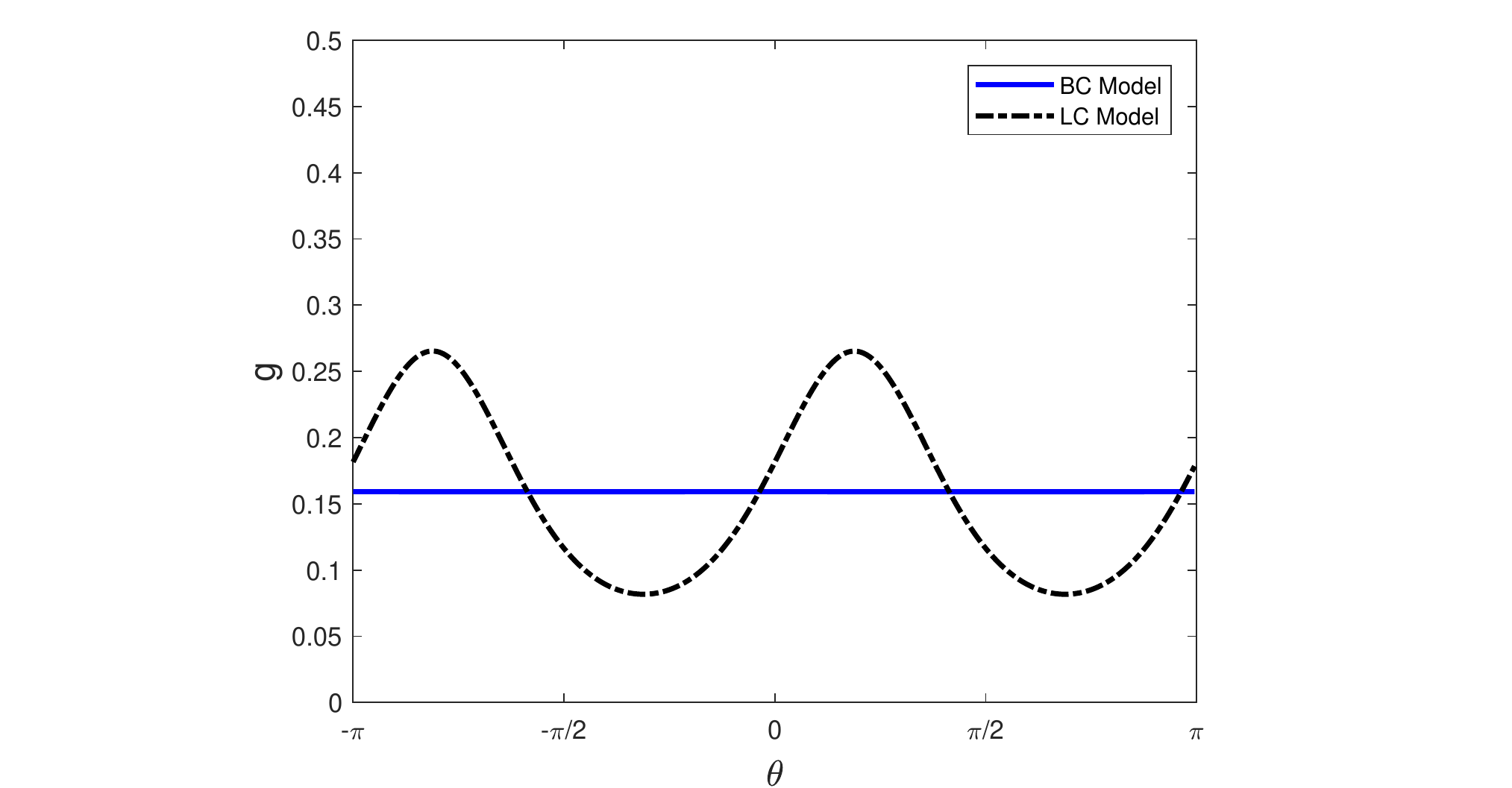}}
\caption{Steady states according to binary collision (BC) model and liquid crystal (LC) model. Figure shows plots of function $g(\theta,t)$ at time $t$ when the solutions reach steady states, according to the numerical simulation of equation \eqref{eq:BCg} 
and the corresponding equation for the liquid crystal model. The solutions start with the same random initial data.    
Left:  $\eta=0.05$, Right:  $\eta=0.2$.
\label{fig:steady_states}
}
\end{figure}

We used the above numercal scheme to obtain solutions of equation \eqref{eq:BCg} and study it's dependence on the diffusivity parameter $\eta=\frac{\mu}{\kappa}.$ For initial data we selected $10^4$ random orientations from the uniform distribution on the interval $[-\pi,\pi].$ The PDF was binned on the mesh in the $\theta$ variable, with bin width $\Delta \theta.$ Figure \ref{fig:cuts}
shows the result of one such simulation in the unstable regime, when the solution converges to a non-uniform steady state (the plot corresponding to $t=25$).  We preformed the comparison of numerical solutions of \eqref{eq:BCg} with the numerical solutions of the liquid crystal model. Figure \ref{fig:steady_states} shows the difference in steady states obtained from both models starting from the same initial data for two different levels of noise. For lower noise levels, the BC solution reaches a less ordered state due to the presence of the factor $|\sin(\theta_1-\theta)|$ in the rate alignment, which decreases the alignment strength among rods. Increasing the noise to the regime where the binary collision model is completely disordered but the liquid crystal model is not, we see the latter is perturbed from the uniform state in the direction of the mode $\sin(2(\theta-\theta_0))$ for some phase $\theta_0$.

\begin{figure}[htbp]
\centering
\includegraphics[clip, trim=0.5cm 8.2cm 0.5cm 9cm, width=1.00\textwidth]{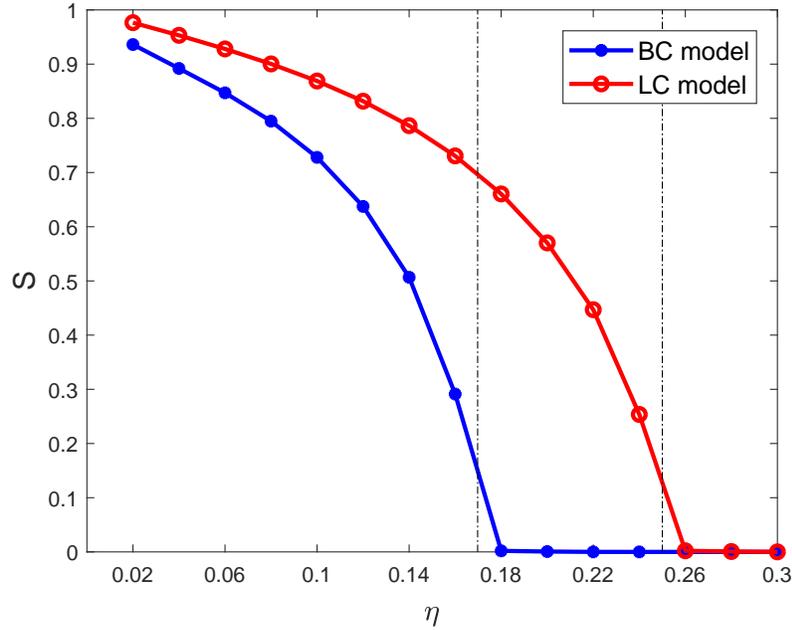}

\caption{Nematic order parameter $S$ as function of diffusivity $\eta.$  Figure shows nematic order $S$ at the steady states, computed numerically, averaged over 20 different simulations with random initial data, when $\eta$ ranges from 0.02 to 0.3 with an increment of 0.02.       Dotted lines mark the critical diffusivites for the corresponding linear models.
\label{fig:phase1}
}
\end{figure}

Figure \ref{fig:phase1} shows the phase transition in the nematic order parameter $S(t) = |{\bf S}(t)|,$ where
\begin{equation}
{\bf S}(t){}={}\int_{-\pi}^\pi e^{2\i\theta} g(\theta,t)\,d\theta,
\end{equation}
is a function of  diffusivity coefficient $\eta =\frac{\mu}{\kappa}$ for both the BC and the LC models.
In these figures, $S$ was computed when the solution reached the steady state, and the average was taken over 1000 distinct solutions with the initial data corresponding to  $10^4$ rods with orientations randomly selected uniformly in the interval $[-\pi,\pi].$  In the ordered regime, the nematic order is stronger in LC model for the same value of $\eta$, while for both models the nematic order decreases at approximately the same rate at the phase transition.

\section{Discussion}

We have developed an approach to modeling interactions of self-propelled rods from a set of microscopic collision rules. Using a Boltzmann-like framework \cite{CercignaniBook}, we derived a mean-field model under the assumption that binary collisions are the dominant type of interaction, and that a collision result in a small change in a rod's orientation. We show for a common choice of the alignment function that the resulting Fokker-Planck equation can be matched to commonly-used mean-field models assuming alignment of rods in a local neighborhood \cite{Peruani2008}, albeit with a modified alignment function. These models have an additional second-order term from added rotational diffusion. Restricting to spatially homogeneous solutions, we calculate both analytically and numerically when the constant solution with a uniform distribution in orientation space becomes unstable, resulting in the emergence of nematic order.

Our approach differs from standard treatments for this type of phenomenon \cite{Peruani2008,Bertin2006,Ginelli2010,Chate2006}. We avoid as much as possible postulating phenomenological rules for how rods align, instead using first principles to derive the form of the mean nematic current, which is a purely local quantity spatially. In particular, our nematic current has an additional factor of $|\sin(\theta_2 - \theta_1)|$ that takes into account the geometry of collisions. In particular, it accounts for a decrease in the collision rate between rods with similar nematic orientations $\theta_1$ and $\theta_2$, as those rods are closer to being parallel with each other and less likely to collide. Furthermore, we show that this additional factor does not change the functional form of the phase transition measured via the nematic order parameter $S$, but does notably raise the densities and lower rotational noise levels at which the transition occurs. The physical interpretation is that a decrease in collisions between rods with similar nematic orientation will decrease the total rate of alignment in the population and make it more susceptible to noise. As such, a higher density or lower rotational diffusion is needed to produce the phase transition to an ordered state.

As a final note on our modeling approach, our treatment of $\delta$ as a small parameter from rescaling $\phi$ is not needed under conditions where rods are locally aligned. The function $\phi$ will naturally be small in such situations assuming the change in orientation on collision is small for small differences in angles, as a large discontinuity at $\phi(0)$ would imply rods actively repel each other when nearly aligned. This suggests our model can be used to analyze situations where the change is angle per collision is not necessarily small, but rods are locally aligned into streams or clusters.

The alignment of active particles can be generated by two general mechanisms. The first is interactions at a distance. These can be through hydrodynamic interactions, e.g. alignment through long range interactions via forces in a medium \cite{riedel05,kumar14,elgeti15}, or through more general flocking interactions \cite{vicsek95,toner98}. The second is short-ranged physical interactions, such as the type we considered in this paper. They depend on the physical geometry of the particles under consideration, arise from physical collisions, and often appear when modeling granular media \cite{narayan07,kudrolli08} or biological phenomena. These phenomena range in scale, but are generally found on either the macromolecular scale when considering interaction of biological filaments \cite{Ndlec97,Schaller10,Sumino12} or cellular populations \cite{Kaiser03,Curtis07,Balagam15b,Li17b,Dell2018,Be'er20}. At larger scales, organisms' sensory organs tend to result in longer-ranged interactions.

The effect the presence our geometric factor has on the influence of level of noise has some interesting implications for active nematics. Since it lowers the threshold at which nematic order is destroyed, this suggesting that systems that depend on purely physical collisions to align are less robust to effects that introduce noise into the rods' orientations. In the context of cellular interactions, this could influence emergent behaviors such as swarming \cite{Be'er20,Kearns03} or aggregation \cite{kiskowski04,Curtis07,Peruani12b} that often depend on some manner of cellular alignment. The noise itself could come externally such as from heterogeneity or external forces in the medium the cells are interacting with, or from internal noise coming from the biochemistry of the cell. In either case, having an alternative alignment mechanism that acts at at longer range, such as through hydrodynamic forces or by actively shaping the environment through the use of an extracellular matrix or biofilm \cite{Wolgemuth02a,Balagam15b,Li17b}, provides a greater level of robustness for when alignment is desired.

\section{Acknowledgments}
Research is supported by National Science Foundation Division of Mathematical Sciences awards 1903270 (to M.P. and I.T.) and 1903275 (to OAI and PM).
We also would like to thank prof. D. Kuzmin (TU Dortmund) for helpful discussions on the numerical methods for the kinetic model presented in this paper.

\bibliographystyle{plain}

\bibliography{references}

\newpage

\begin{figure}[htbp]
\centering
\includegraphics[clip, trim=0.5cm 8.cm 0.5cm 9cm, width=1.00\textwidth]{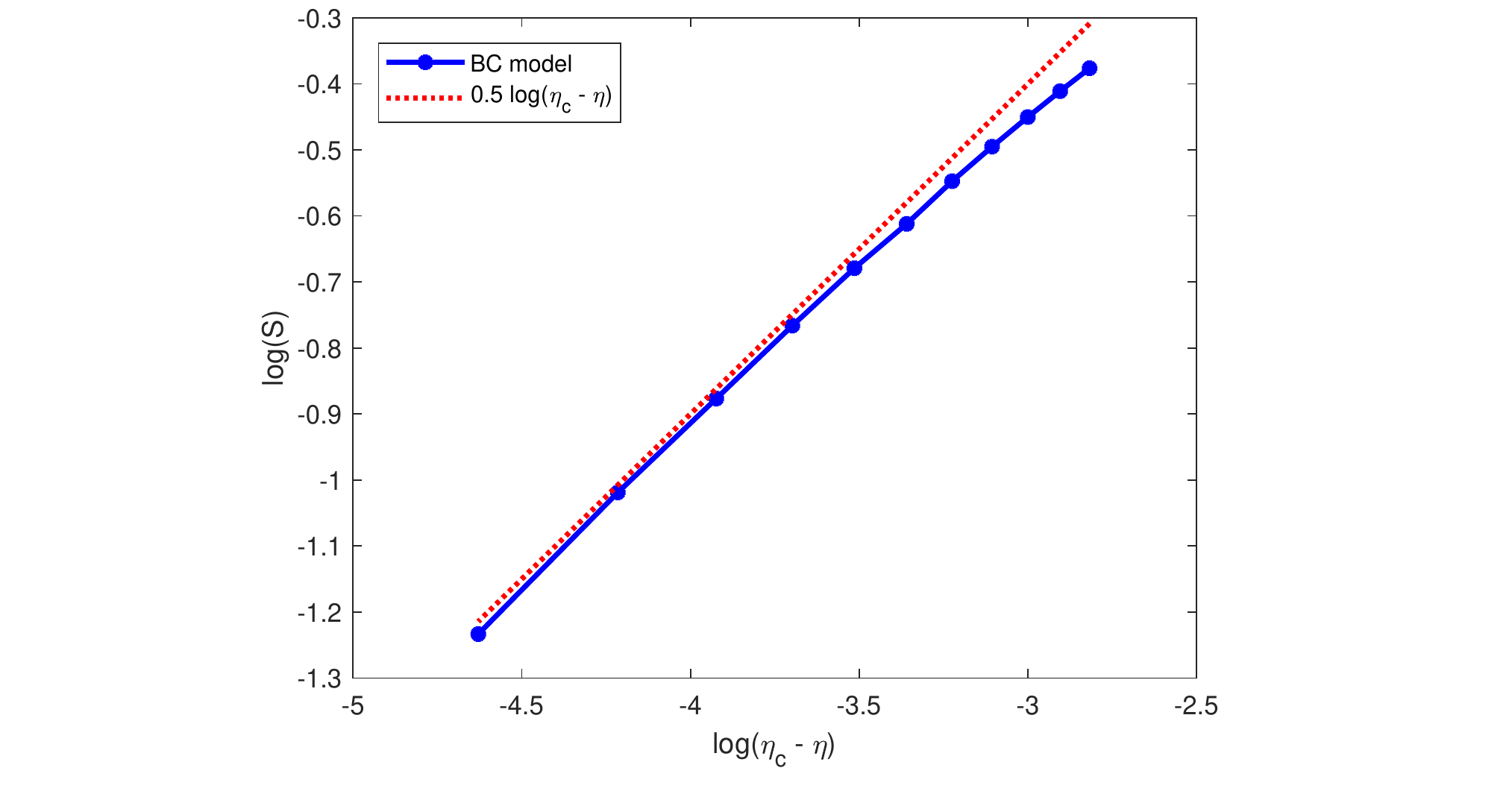}

\caption{Rate of decay of the nematic order parameter $S$ as function of diffusivity $\eta$ for $\eta < \eta_c$.  %Same as Figure \ref{fig:rate1}, but plotted on the log-log scale.
Red dashed line corresponds to the fit $\log(S) \sim 0.5 \log(\eta_c - \eta)$. Note that $\log(\eta_c - \eta) \approx -4.5$ corresponds to $\eta=1.6$.
\label{fig:rate2}
}
\end{figure}

\end{document}